\documentclass[a4paper]{jpconf}
\usepackage{graphicx}
\usepackage{psfrag}
\begin{document}
\title{Oscillations in Procyon A \\ First results from a multi-site campaign}

\author{S~Hekker$^1$, T~Arentoft$^2$, H~Kjeldsen$^2$, T~R~Bedding$^3$, J~Christensen-Dalsgaard$^2$, S~Reffert$^4$, H~Bruntt$^3$, R~P~Butler$^5$, L~L~Kiss$^3$, S~J~O'Toole$^3$, E~Kambe$^6$, H~Ando$^7$, H~Izumiura$^6$, B~Sato$^8$, M~Hartmann$^9$, A~P~Hatzes$^9$, T~Appourchaux$^{10}$, C~Barban$^{11}$, G~Berthomieu$^{12}$, F~Bouchy$^{13}$, R~A~Garc\'{\i}a$^{14}$, J-C~Lebrun$^{15}$, M~Marti\'c$^{15}$, E~Michel$^{11}$, B~Mosser$^{11}$, P~A~P~Nghiem$^{14}$, J~Provost$^{12}$, R~Samadi$^{11}$, F~Th\'evenin$^{12}$, S Turck-Chi\`eze$^{14}$, S~A~Bonanno$^{16}$, S~Benatti$^{17}$, R~U~Claudi$^{17}$, R~Cosentino$^{18}$, S~Leccia$^{19}$, S~Frandsen$^{2}$, K~Brogaard$^{2}$, F~Grundahl$^{2}$, H~C~Stempels$^{20}$, M~Bazot$^{2}$, T~H~Dall$^{21}$, C~Karoff$^{2}$, F~Carrier$^{22}$, P~Eggenberger$^{23}$, D~Sosnowska$^{24}$, R~A~Wittenmyer$^{25}$, M~Endl$^{26}$ and T~S~Metcalfe$^{27}$}

\address{$^1$ Leiden Observatory, Leiden University, P.O. Box 9513, 2300 RA Leiden, The Netherlands}
\address{$^2$ Department of Physics and Astronomy, University of Aarhus, DK-8000 Aarhus C, Denmark}
\address{$^3$ School of Physics A28, University of Sydney, NSW 2006, Australia}
\address{$^4$ ZAH, Landessternwarte Heidelberg, K\"onigstuhl 12, D-69117 Heidelberg, Germany}
\address{$^5$ Carnegie Institute of Washington, Department of Terrestrial Magnetism, 5241 Broad Branch Road NW, Washington, DC 20015-1305, USA}
\address{$^6$ Okayama Astrophysical Observatory, National Astronomical Observatory of Japan, Asakuchi, Okayama 719-0232, Japan}
\address{$^7$ National Astronomical Observatory of Japan, Mitaka, Tokyo 181-8588, Japan}
\address{$^8$ Tokyo Institute of Technology, 2-12-1-S6-6, Okayama, Meguro-ku, Tokyo 152-8550, Japan}
\address{$^9$ Th\"uringer Landessternwarte Tautenburg, Sternwarte 5, 07778 Tautenburg, Germany}
\address{$^{10}$ Institut d'Astrophysique Spatiale, Universit\'e Paris XI-CNRS, B\^atiment 121, 91405 Orsay cedex, France}
\address{$^{11}$ LESIA, CNRS, Universit\'e Pierre et Marie Curie, Universit\'e Denis Diderot, Observatoire de Paris, 92195 Meudon cedex, France}
\address{$^{12}$ Laboratoire Cassiop\'ee, UMT CNRS 6202, Observatoire de la C\^ote d'Azur, BP 4229, 06304 Nice cedex, France}
\address{$^{13}$ Institut d'Astrophysique de Paris, CNRS, 98$^{bis}$ BD Arago, 75014 Paris, France}
\address{$^{14}$ DAPNIA/DSM/Service d'Astrophysique, CEA/Saclay, 91191 Gif-sur Yvette cedex, France}
\address{$^{15}$ Service d'A\'eronomie du CNRS, BP 3, 91371 Verri\`eres le Buisson, France}
\address{$^{16}$ INAF- Osservatorio Astrofisico di Catania, via S. Sofia 78, 95123 Catania, Italy}
\address{$^{17}$ INAF- Osservatorio Astronomica di Padova, Vicolo Osservatorio 5 35122 Padova, Italy}
\address{$^{18}$ INAF-Telescopio Nazionale Galileo, 38700 Santa Cruz de La Palma, TF, Spain}
\address{$^{19}$ INAF-Astronomical Observatory of Capodimonte, Salita Moiariello 16, I-80131 Napoli, Italy}
\address{$^{20}$ School of Physics and Astronomy, University of St Andrews, North Haugh, St Andrews KY16 9SS, Scotland, UK}
\address{$^{21}$ Gemini Observatory, 670 N. A'ohoku Pl., Hilo, HI 96720, USA}
\address{$^{22}$ Instituut voor Sterrenkunde, KU Leuven, Celestijnenlaan 200D, 3001 Leuven, Belgium}
\address{$^{23}$ Institut d'Astrophysique et de G\'eophysique de l'Universit\'e de Li\`ege, All\'ee du 6 Ao\^ut, 4000 Li\`ege, Belgium}
\address{$^{24}$ Observatoire de Gen\`eve, Universit\'e de Gen\`eve, 51 chemin des Maillettes, 1290 Sauverny, Switzerland}
\address{$^{25}$ Astronomy Department, The University of Texas at Austin, Austin, TX 78712, USA}
\address{$^{26}$ McDonald Observatory, The University of Texas at Austin, Austin, TX 78712, USA}
\address{$^{27}$ High Altitude Observatory and Scientific Computing Division, NCAR, PO Box 3000, CO 80307, USA}

\ead{saskia@strw.leidenuniv.nl}

\begin{abstract}
Procyon A is a bright F5IV star in a binary system. Although the distance, mass and angular diameter of this star are all known with high precision, the exact evolutionary state is still unclear. Evolutionary tracks with different ages and different mass fractions of hydrogen in the core pass, within the errors, through the observed position of Procyon A in the Hertzsprung-Russell diagram. For more than 15 years several different groups have studied the solar-like oscillations in Procyon A to determine its evolutionary state. Although several studies independently detected power excess in the periodogram, there is no agreement on the actual oscillation frequencies yet. This is probably due to either insufficient high-quality data (i.e., aliasing) or due to intrinsic properties of the star (i.e., short mode lifetimes). Now a spectroscopic multi-site campaign using 10 telescopes world-wide (minimizing aliasing effects) with a total time span of nearly 4 weeks (increase the frequency resolution) is performed to identify frequencies in this star and finally determine its properties and evolutionary state. 
\end{abstract}

\section{Introduction}
The bright F5 subgiant Procyon A is the primary of  an astrometric binary system with a white dwarf in a 40 year orbit. Procyon A is the brightest northern-hemisphere asteroseismology candidate with well-determined characteristics, such as distance, mass and angular diameter. 
Brown {\etal} \cite{brown1991} were among the first to observe an excess power between 0.5 and 1.5 mHz in radial velocity observations confirmed by several other radial velocity studies, e.g. \cite{martic1999}, \cite{bouchy2002}, \cite{kambe2003}, \cite{martic2004}, \cite{eggenberger2004}, \cite{claudi2005}, \cite{leccia2007}. So far these studies have independently revealed detections of power excess, but there is no agreement yet on the actual oscillation frequencies. This may be due to aliases present in the spectral window, short mode lifetimes, shifts from the asymptotic relation due to avoided crossings, or any combination of these factors. Although the frequencies are not yet known in detail, most studies obtain a large frequency spacing of $55 \pm 1$ $\mu$Hz.

\begin{table}[h!]
\caption{\label{stelpar} Stellar parameters of Procyon A from \cite{allende2002}.}
\begin{center}
\begin{tabular}{lr@{$\pm$}l}
\br
Mass [M$_{\odot}$]& 1.42 & 0.06\\
T$_{\rm eff}$ [K] & 6512 & 49 \\
Radius [R$_{\odot}$] & 2.071 & 0.020\\
$\rm[Fe/H]$  [dex] & $-$0.09 & 0.03 \\
$v_{\rm rot}\sin i$ [km\,s$^{-1}$] & 3.16 & 0.50\\
\br
\end{tabular}
\end{center}
\end{table}

Another point of discussion is the fact that \cite{matthews2004} did not detect any power excess in their MOST photometry and published Procyon A as a flat liner. Other photometric studies such as WIRE \cite{bruntt2005} and a reanalysis of the MOST 2004 data \cite{regulo2005} claim to detect power excess in the same region as the radial velocity studies. A recent reanalysis of the MOST 2004 data and the analysis of new MOST data taken in 2005 reinforce the null detection of p-modes \cite{guenther2007}. This issue is discussed more extensively by \cite{bedding2005}, who claim that the non-detection of oscillations in Procyon by the MOST satellite is fully consistent with the ground based radial-velocity studies due to a combination of several noise sources and the low photometric amplitude of the oscillations.

Procyon A is in a very interesting evolutionary state near the end of its main sequence life. The stellar parameters (see Table~\ref{stelpar}) are all known to high precision, but several evolutionary tracks with different ages and different hydrogen core mass fractions overlap, within the errors, with its position in the HR-diagram. The exact evolutionary state of a star can be revealed by means of asteroseismology and therefore the oscillation frequencies are needed.

To determine the actual frequencies of Procyon A a ground-based multi-site campaign using 10 telescopes with a total time span of nearly 4 weeks was performed from December 28, 2006 till January 22, 2007. Here we present first results of this campaign. In Section 2 the campaign is described, while Section 3 discusses the way the data of the different telescopes are combined, and a final power spectrum is obtained. Some concluding remarks and future prospects are provided in Section 4.

\section{Multi-site campaign}
For this multi-site campaign 10 telescopes around the world with high-resolution spectrographs were used, and in total 20899 spectra were obtained during 472 hours. During the central 9 days of this campaign the coverage was 89\%.

For most of the facilities the available reduction pipelines were used to compute the radial velocity variations. For the SOPHIE spectrograph mounted on the 1.93 m telescope, Observatoire de Haute Provence, France this was one of the first runs after commissioning. The capabilities of SOPHIE for seismology are shown by \cite{mosser2007} and the iodine method used at Okayama is described by \cite{kambe2007} both using data from the current multi-site campaign. For the FIES spectrograph mounted on the Nordic Optical Telescope, La Palma the observations of Procyon were part of the commissioning run and the reduction pipeline was optimised using the Procyon data.

In Table~\ref{obs} we list the time span and the number of hours of observations, the number of spectra collected as well as the mean uncertainty for the data obtained at each telescope. 

\begin{table}
\caption{\label{obs} Observatory, telescope, spectrograph, start of run (beginning of the night, local time), end of run (end of the night, local time), number of observed hours, number of spectra, mean uncertainty on the spectra ($\sigma$) in m\,s$^{-1}$ for all facilities used for this multi-site campaign (ordered according to starting date and length of the run).}
\begin{center}
\begin{tabular}{lllllrrc}
\br
Observatory & Telescope & Spectrograph & Start & End & Hours & Spectra & $\sigma$ [m\,s$^{-1}$]\\
\mr
Okayama & 1.9 m & HIDES & Dec 28 & Jan 17 & 90 & 1997 & 2.0\\
McDonald & 2.7 m & & Dec 28 & Jan 3 & 15 & 1719 & 4.6\\
TLS & 2.0 m & & Dec 29 & Jan 18 & 19 & 494 & 2.9\\
Siding Spring & 3.9 m AAT & UCLES & Dec 29 & Jan 10 & 53 & 2451 & 1.7\\
La Palma & 2.5 m NOT & FIES & Dec 31 & Jan 10 &17 & 1087 & 5.2\\
OHP &1.9 m & SOPHIE & Jan 2 & Jan 11 & 57 & 3924 & 1.9\\
La Silla & 3.6 m & HARPS & Jan 3 & Jan 11 & 62 & 5698 & 0.8\\
La Palma & 3.6 m TNG & SARG & Jan 8 & Jan 12 & 19 & 693 & 1.9\\
Lick & 0.6 m CAT & Hamilton &  Jan 7 & Jan 22 & 106 & 1900 & 2.7\\
La Silla & 1.2 m & CORALIE & Jan 11 & Jan 17 & 34 & 936 & 2.0\\
\br
\end{tabular}
\end{center}
\end{table}

\section{Data combination}
The radial velocities from all different telescopes are combined using the method developed by \cite{butler2004} for $\alpha$ Cen A. For each spectrum the measurement uncertainty $\sigma_{i}$ is used to assign a weight ($w_{i}=1/\sigma^{2}_{i}$). This weight is re-scaled to match the actual noise per site and for each night, using the following relation: 
\begin{equation}
\sigma^{2}_{\rm amp} \sum^{N}_{i=1} \sigma^{-2}_{i} = \pi,
\end{equation}
where the mean variance of the data is set to be equal to the variance deduced from the noise level $\sigma_{\rm amp}$ in the amplitude spectrum.
We then follow a procedure very similar to the one described in \cite{butler2004} to identify poor data points, and adjust the uncertainties of these points to give them lower weights. 
With this procedure the combined data set is noise-optimised, i.e., the weights have been chosen to minimize the noise.

\begin{figure}
\begin{center}
\includegraphics[width=\linewidth]{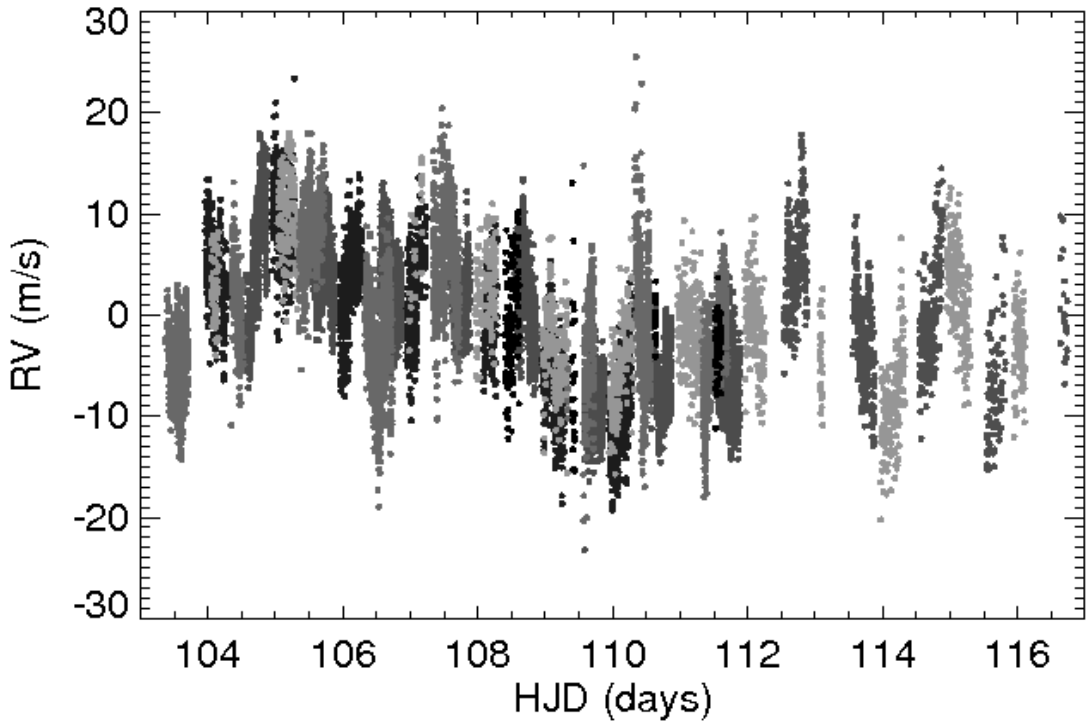}
\caption{\label{centrpart}The central part of the radial velocity observations as a function of time. The different colours indicate data obtained with different telescopes.}
\end{center}
\end{figure}

\begin{figure}
\psfrag{HJD - 54100}[c][c][0.8][0]{{\fontfamily{uarial}\selectfont HJD-2454100}}
\begin{center}
\includegraphics[width=\linewidth]{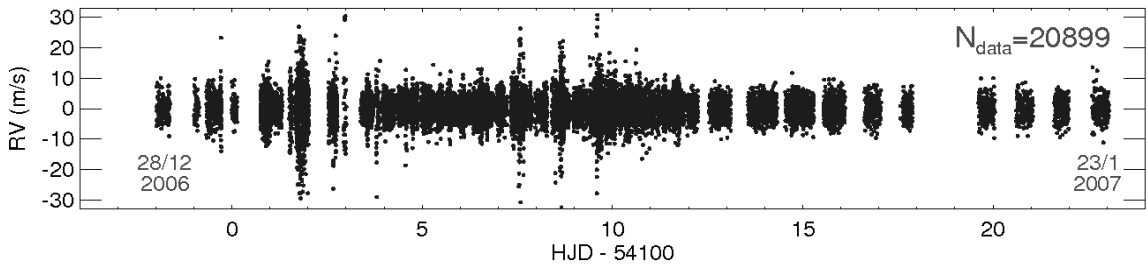}
\caption{\label{alldata}All 20899 radial velocity data points as a function of time. The slow drifts have been subtracted. The errors in the data points are 1--3 m\,s$^{-1}$ and the coverage during the central 9 days is 89\%.}
\end{center}
\end{figure}

In Figure~\ref{centrpart} the central part of the radial velocity data of the campaign is shown. From this figure it becomes clear that there is variation with a periodicity of the order of 6 days, while the solar-like oscillations are in the range of 10--30 minutes. This slow drift is present in the data of all observatories and is therefore most likely intrinsic to the star and not due to instrumental effects. 
In Figure~\ref{alldata} all radial velocity variations obtained with the different telescopes is shown, with the slow drift subtracted.

\begin{figure}
\begin{center}
\begin{minipage}{15pc}
\includegraphics[width=15pc]{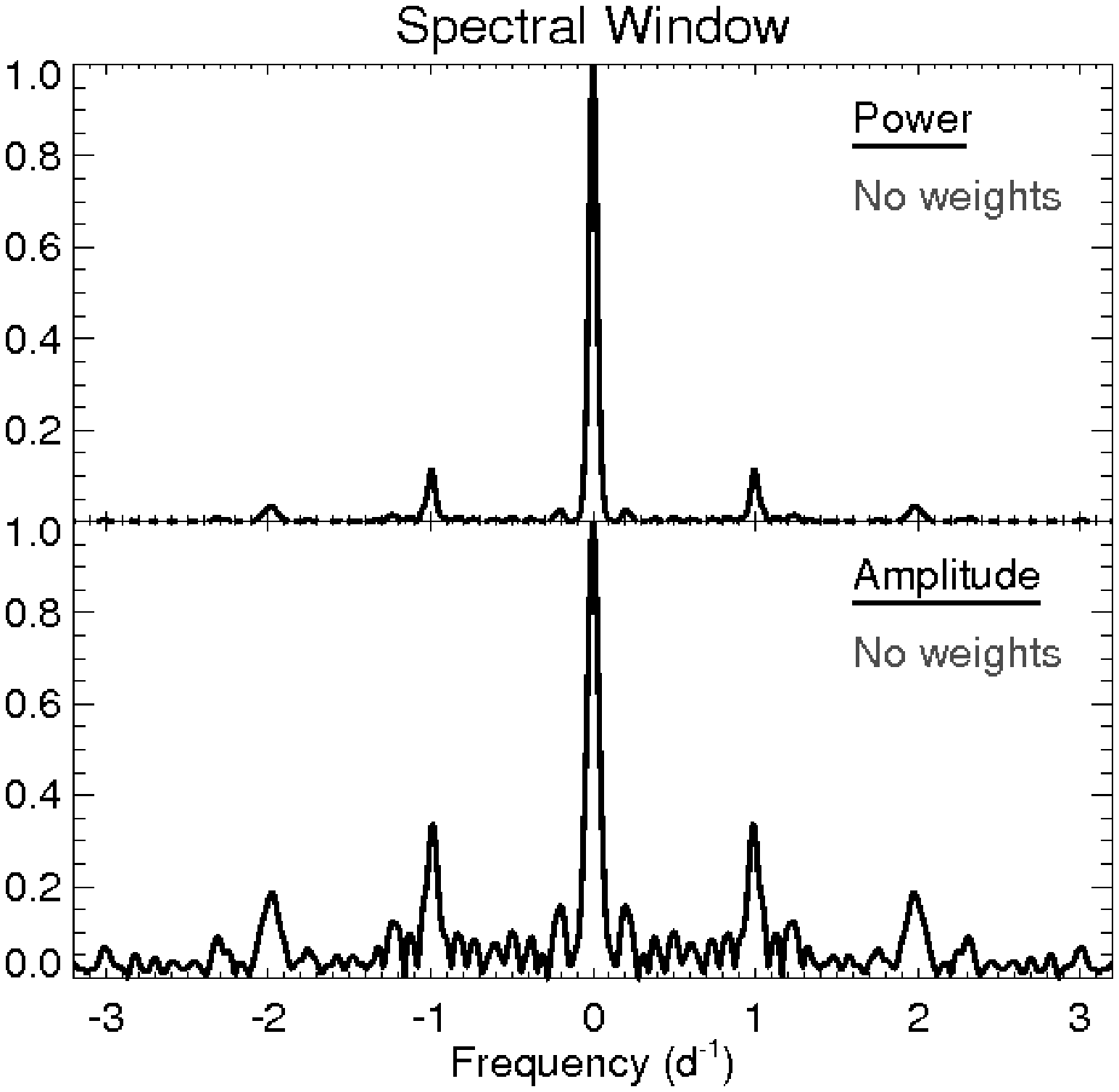}
\end{minipage}\hspace{2pc}
\begin{minipage}{15pc}
\includegraphics[width=15pc]{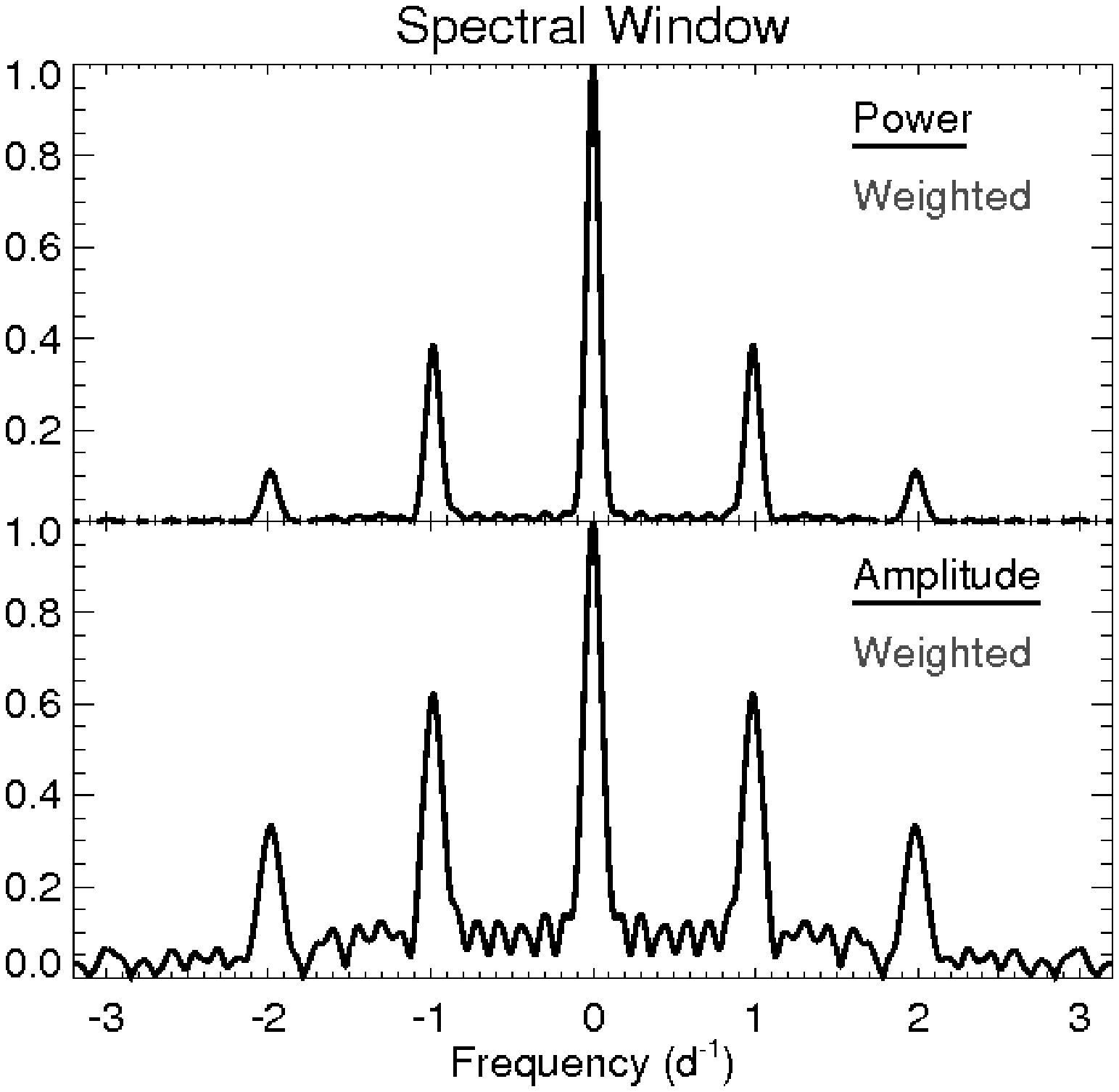}
\end{minipage}\hspace{2pc}
\begin{minipage}{15pc}
\includegraphics[width=15pc]{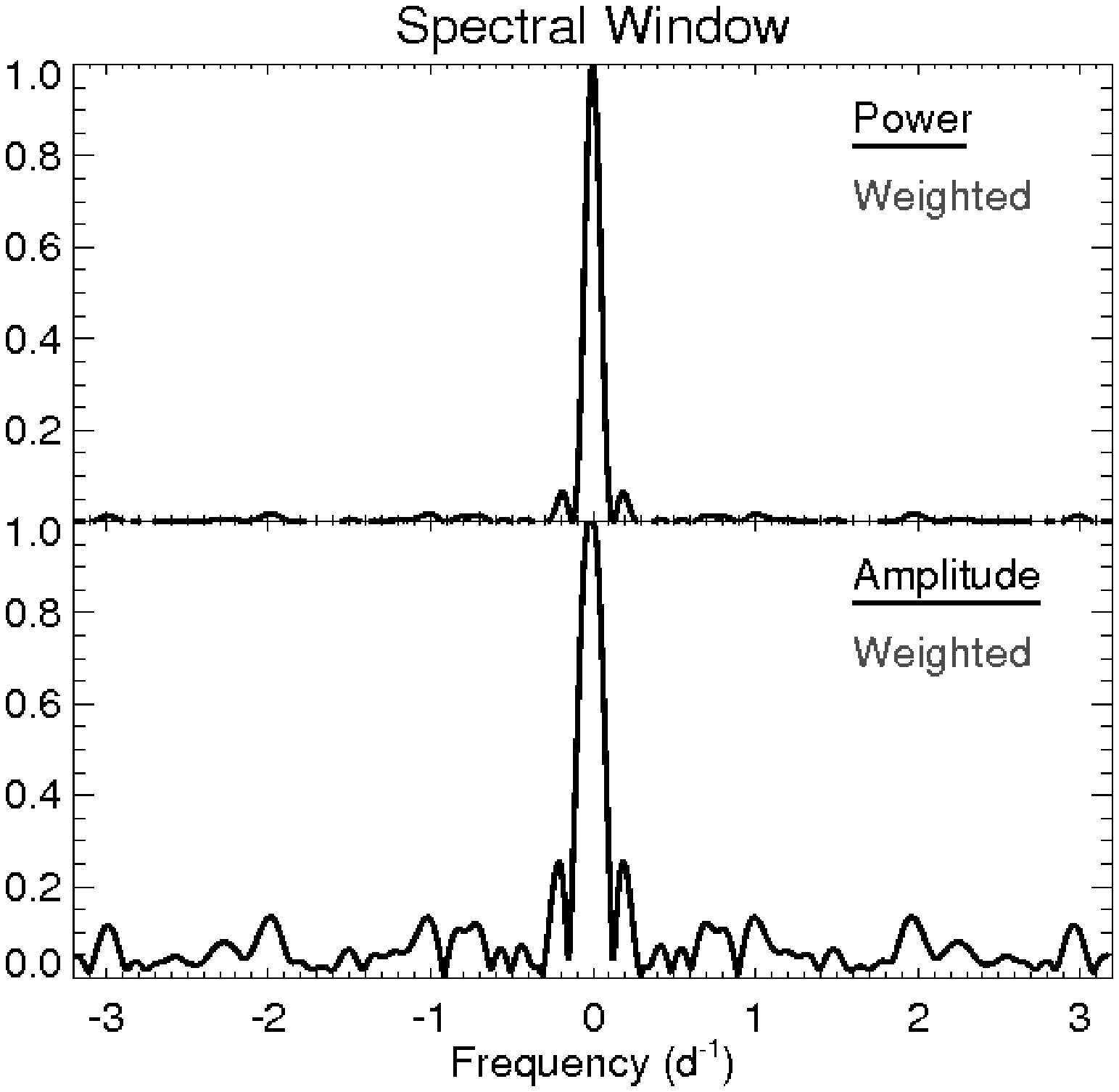}
\end{minipage}\hspace{2pc}
\begin{minipage}{15pc}
\caption{\label{window}Window functions for data without weights (top left), noise-optimised weights (right) and window-optimised weights (lower left).}
\end{minipage}
\end{center}
\end{figure}

The data obtained at the best sites will get the highest weights in a noise-optimised data set. This enhances the sidelobes in the spectral window function. As is well known, these sidelobes complicate the oscillation spectrum, especially for weaker modes, and could lead to mis-identification. In order to optimise the window function, the weights are adjusted on a night-by-night basis, i.e., we allocate an adjustment factor for each night and telescope. The noise-optimised weights are multiplied by these factors and a new spectral window is calculated. This process is iterated to minimize the height of the sidelobes as was also done by \cite{bedding2004}. In this way we create a window-optimised data set for which the sidelobes are effectively negligible.
In Figure~\ref{window}, window functions are shown for the data without weights, with noise-optimised weights and with window-optimised weights.

\begin{figure}
\begin{center}
\includegraphics[width=\linewidth]{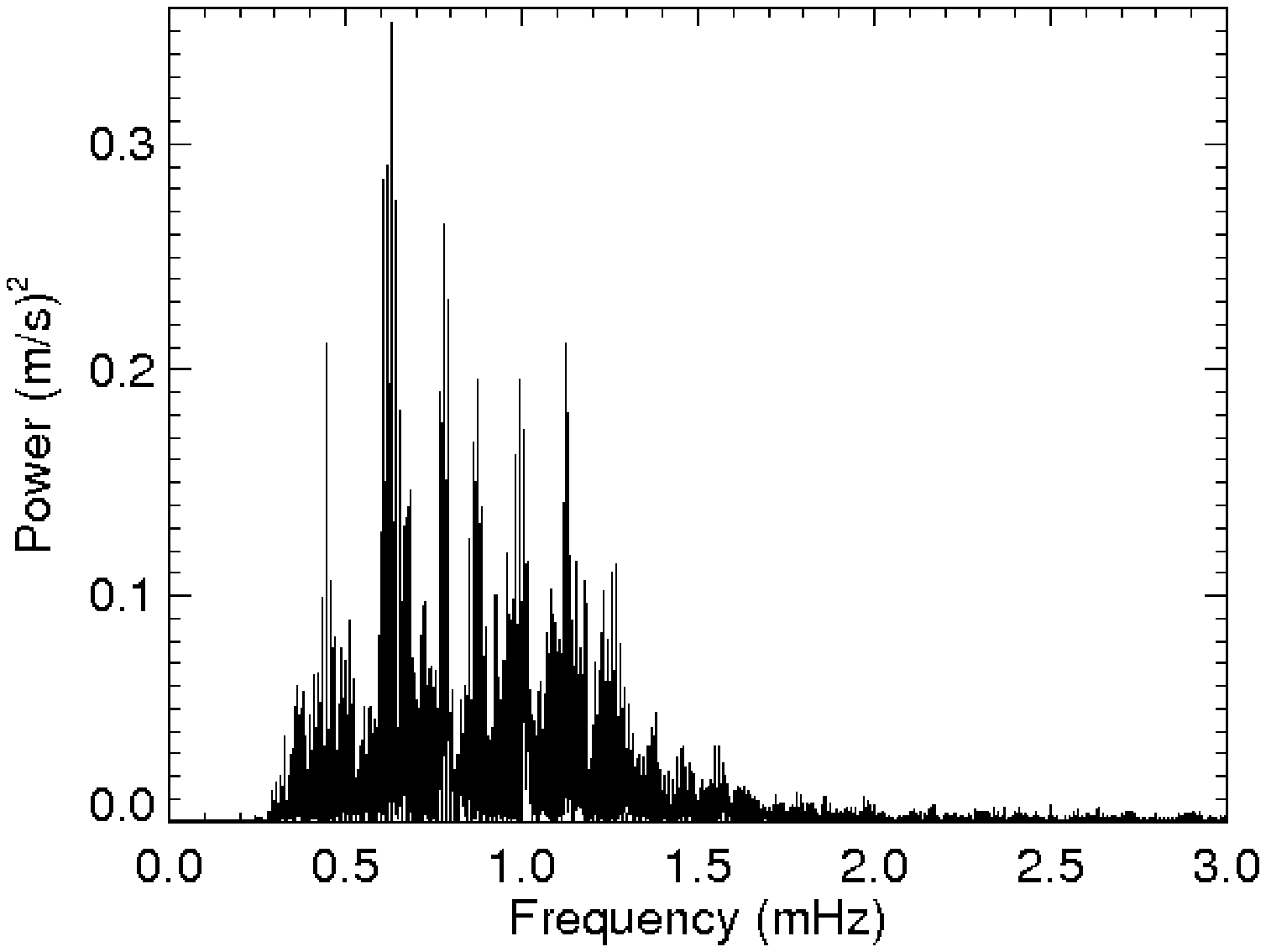}
\caption{\label{power}Final power spectrum of Procyon A with window-optimised weights. The data are high-pass filtered to remove low-frequency drifts.}
\end{center}
\end{figure}

From the combined high-pass filtered data with window-optimised weights the power spectrum is calculated and shown in Figure~\ref{power}. From this power spectrum we can recognise a regular pattern with a large frequency separation of about 55 $\mu$Hz, consistent with earlier determinations.

\section{Concluding remarks and future prospects}
The Procyon campaign presented here is the largest spectroscopic campaign, so far, aimed at the detection and identification of solar-like oscillations. Using 10 telescopes over a time span of nearly 4 weeks provided us with a unique data set with high time coverage and frequency resolution. 
The details of the data processing methods will be fully described by \cite{arentoft2008}, while the oscillation frequencies extracted from the full data set acquired during the spectroscopic Procyon campaign will be presented by \cite{bedding2008}.

\end{document}